\documentclass[reprint,superscriptaddress, amsmath,amssymb,aps,prb]{revtex4-2}

\usepackage{multirow}
\usepackage{amssymb}
\usepackage{amsmath}
\usepackage[dvipdfmx]{graphicx}
\usepackage{dcolumn}
\usepackage{bm}
\usepackage[dvipdfmx]{color}
\usepackage{threeparttable}
\usepackage[dvipdfmx]{hyperref}
\hypersetup{
    colorlinks=true,
    linkcolor=blue,
    filecolor=magenta,      
    urlcolor=cyan,
    pdftitle={manuscript},
    pdfpagemode=FullScreen,
    }

\newcommand{\aI}{$\alpha$-ET$_2$I$_3$}

\newcommand{\I}{$^{127}$I}
\newcommand{\Dl}{D$\rm _L$}
\newcommand{\Dh}{D$\rm _H$}
\newcommand{\nuQcal}{$\nu_{\rm Q}^{\rm cal}$}
\newcommand{\nuQexp}{$\nu_{\rm Q}^{\rm exp}$}

\begin{document}

\preprint{}
\title{Role of hydrogen bonding in charge-ordered organic conductor $\alpha$-(BEDT-TTF)$_2$I$_3$ probed by $^{127}$I nuclear quadrupole resonance}
\author{T.~Kobayashi}
 \email{tkobayashi@mail.saitama-u.ac.jp}
 \affiliation{Graduate School of Science and Engineering, Saitama University, Saitama, 338-8570, Japan}
 \affiliation{Research and Development Bureau, Saitama University, Saitama 338-8570, Japan}
 \affiliation{Meson Science Laboratory, RIKEN, Saitama 351-0198, Japan}
 \author{Y.~Kato}
 \affiliation{Graduate School of Science and Engineering, Saitama University, Saitama, 338-8570, Japan}
\author{H.~Taniguchi}
 \affiliation{Graduate School of Science and Engineering, Saitama University, Saitama, 338-8570, Japan}
\author{T.~Tsumuraya}
 \affiliation{Magnesium Research Center, Kumamoto University, 860-8555, Japan}
\author{K.~Hiraki}
 \affiliation{Center for Integrated Science and Humanities, Fukushima Medical University, Fukushima, 960-1295, Japan}
\author{S.~Fujiyama}
 \affiliation{Meson Science Laboratory, RIKEN, Saitama 351-0198, Japan}
 
\date{\today}

\begin{abstract}
We present $^{127}$I nuclear quadrupole resonance spectra and nuclear relaxation of $\alpha$-(BEDT-TTF)$_2$I$_3$ that undergoes a charge-ordering transition. Only one of the two I$_3$ anion sites shows a significant differentiation in the electric field gradients across the first-order transition. The charge modulation only in the BEDT-TTF layers can not reproduce; instead, an anion-donor interaction accompanied by hydrogen bonding is necessary. The dominating source for the nuclear relaxation is the local libration of the I$_3$ anions, but an anomalous peak is detected just below the transition, as observed by $^{13}$C NMR. 
\end{abstract}

\maketitle

\section{Introduction}
Electronic correlation in a quasi-two-dimensional organic conductor (BEDT-TTF)$_{2}X$ stabilizes various ground states, including quantum spin liquids or Dirac electron systems. Molecular arrangements of the BEDT-TTF [bis(ethylenedithio)tetrathiafulvalene, hereafter abbreviated as ET] in the alternating conducting planes specify the electronic correlation leading to various physical properties~\cite{Seo2000,Miyagawa2000}. 
Of these, $\alpha$, $\beta''$, or $\theta$ type molecular arrangements establish two-dimensional 3/4-filled (1/4-filled hole) electronic bands. Long-range Coulomb interactions cause ground states such as charge ordering (CO), ferroelectricity~\cite{Yamamoto2008,Tomic2015}, charge glass~\cite{Kagawa2013}, and superconductivity~\cite{Mori1998,Merino2001,Bangura2005,Pustogow2019}.

\begin{figure}[tbp]
\begin{center}
\includegraphics[width=8cm]{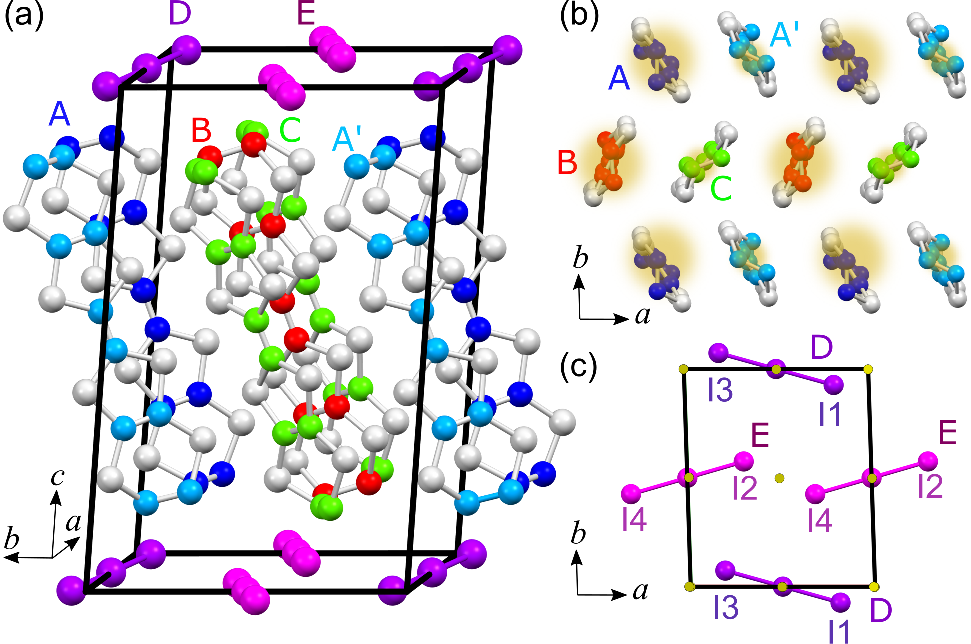}
\end{center}
\caption{Crystal structure of (a) $\alpha$-I$_3$~\cite{Kitou2021}. BEDT-TTF molecules labeled as A, B (A$'$, C) show charge-rich (charge-poor) sites for $T<T_\mathrm{CO}$. (b) Top view of the conduction plane in the CO state. (c) Insulating I$_3$ plane. The inversion center is lost for $T<T_\mathrm{CO}$, by which I1 and I3, and I2 and I4 sites become crystallographically inequivalent.}
\label{fig1}
\end{figure}

A quasi-two-dimensional organic conductor $\alpha$-ET$_2$I$_3$, shown in Fig.~\ref{fig1}(a), is one of the most studied organic conductors~\cite{Dressel2020}. 
This material has a quarter-filled electronic band, showing a Mott-insulating instability. 
One prominent property of $\alpha$-ET$_2$I$_3$ is the realization of a Dirac electron system under applied pressure~\cite{Tajima2000,Katayama2006,Kajita2014,Hirata2017,Fujiyama2022}. 
At ambient pressure, the system shows a metal-insulator transition at $T_\mathrm{CO} \approx 135$~K associated with the CO transition~\cite{Bender1984,Yamamoto2008,Tomic2015}. 
It is widely accepted that the modulation pattern of the CO is not checkerboardlike but zigzag-chain-like with charge-rich sites A and B, and charge-poor sites A$'$ and C [Fig.~\ref{fig1}(b)], as revealed by x-ray structural analysis \cite{Kakiuchi2007}, infrared and Raman spectroscopies \cite{Wojciechowski2003,Ivek2011,Yakushi2012}, and $^{13}$C-NMR spectroscopy~\cite{Takano2001,Hirose2010,Hirata2011,Ishikawa2016}.

%%%%%%%%%%%%%%%%%%%%%%%%%%% FIG1 %%%%%%%%%%%%%%%%%%%%%%%%%%%%%%%%%%%
%%%%%%%%%%%%%%%%%%%%%%%%%%% FIG1 %%%%%%%%%%%%%%%%%%%%%%%%%%%%%%%%%%%

Despite the consensus on the modulation pattern, issues remain regarding the mechanism of the CO transition. 
Many theoretical studies have been conducted and a variety of proposals have been made. 
Coulomb repulsion in the quarter-filled band of the ET layers has been proposed~\cite{Kino1995,Kino1996,Seo1997,Seo2000,Tanaka2016}, followed by a claim that the electron-phonon interaction plays a crucial role to reproduce the CO pattern~\cite{Clay2002,Tanaka2008,Miyashita2010,Udagawa2007}. 
While these theories consider only interactions within the intra-ET layers, some approaches actively incorporate the role of the anion layer through hydrogen bonding. 
Reference~[\onlinecite{Alemany2012}] pointed out the significant role of the H-I$^{-1/2}$ $\sigma$ bond between the anion and edge protons in the ET molecule. 
Since the energy difference between the $\sigma$ bond and $\pi$ electrons of donor molecules are far apart, the $\sigma$ hybridization is expected to conserve the total charge density in the ET layers. 
However, the $\sigma$ bond can constrain atomic positions across the structural phase transition, and also the problem regarding the stability of the Mottness of the quarter-filled band perturbed by the hydrogen bond attracts much attention. 

In this paper, we propose $^{127}$I-nuclear quadrupole resonance (NQR) as a microscopic tool to examine the specific H-I$^{-1/2}$ bonds. 
The NQR spectra of \aI\ are qualitatively reproduced using ﬁrst-principles calculations, which enables us to claim the microscopic interactions that occur during the CO transition. 
Utilizing another merit of NQR, unlike with $^{13}$C NMR, we can directly observe static and dynamic charge properties via an electric field gradient (EFG), and we found the zigzaglike charge modulation in the CO state across a first-order transition. 
Only one of two I-spectral lines shows significant splitting at $T_\mathrm{CO}$, which cannot be reproduced by an electrostatic interaction only from the ET layers, and an interaction between the ET and I$_3$ layers through such a hydrogen bonding is necessary. 
The nuclear spin-lattice relaxation rate ($1/T_{1}$) is dominated by local fluctuation of the I$_3$ anions, of which the energy is five times smaller than the Debye frequency. 
We also found enhancements of $1/T_{1}$ just below $T_\mathrm{CO}$.

\section{Experiments}
Single crystals of \aI\ were grown by the standard electrochemical reaction \cite{Bender1984}.
The $^{127}$I-NQR experiment was performed on a polycrystalline sample of $14.2$~mg under zero magnetic fields.
The NQR spectrum was obtained by fast Fourier transformation of the spin-echo signal with a $\pi/2$--$\pi$ pulse sequence, where the typical $\pi/2$ pulse length was $2.5$~$\mu$s. 
The $1/T_1$ were measured using the inversion and saturation recovery methods and were obtained by fitting the magnetization curves using 
\begin{equation}
    1-\frac{M(t)}{M(\infty)} \propto \frac{3}{28}\exp{\left(-\frac{3t}{T_1}\right)} + \frac{25}{28}\exp{\left(-\frac{10t}{T_1}\right)} 
\end{equation}
(see the Supplemental Material \cite{Supplement} and also Refs.~\cite{MacLaughlin1971,Daniel1964} therein).
Here, $t$ is the interval between inversion (saturation) and the first $\pi/2$ pulses, and $M(t)$ is the magnetization at time $t$. 
We confirmed that the same $T_1$ values were obtained by both methods in the appropriate temperature range. 
EFGs at the iodine sites of \aI\ were calculated using the first-principles calculations based on the full-potential linearized augmented plane-wave (FLAPW) method~\cite{Wimmer1981, Yu1991}. 
The calculations are based on density-functional theory (DFT), and the exchange and correlation potential is represented by a generalized gradient approximation in the Perdew-Burke-Ernzerhof formula~\cite{Supplement,Perdew1996}.

\section{Results and discussion}
\subsection{NQR spectra}
\begin{figure}[tbp]
\begin{center}
\includegraphics[width=8.5cm]{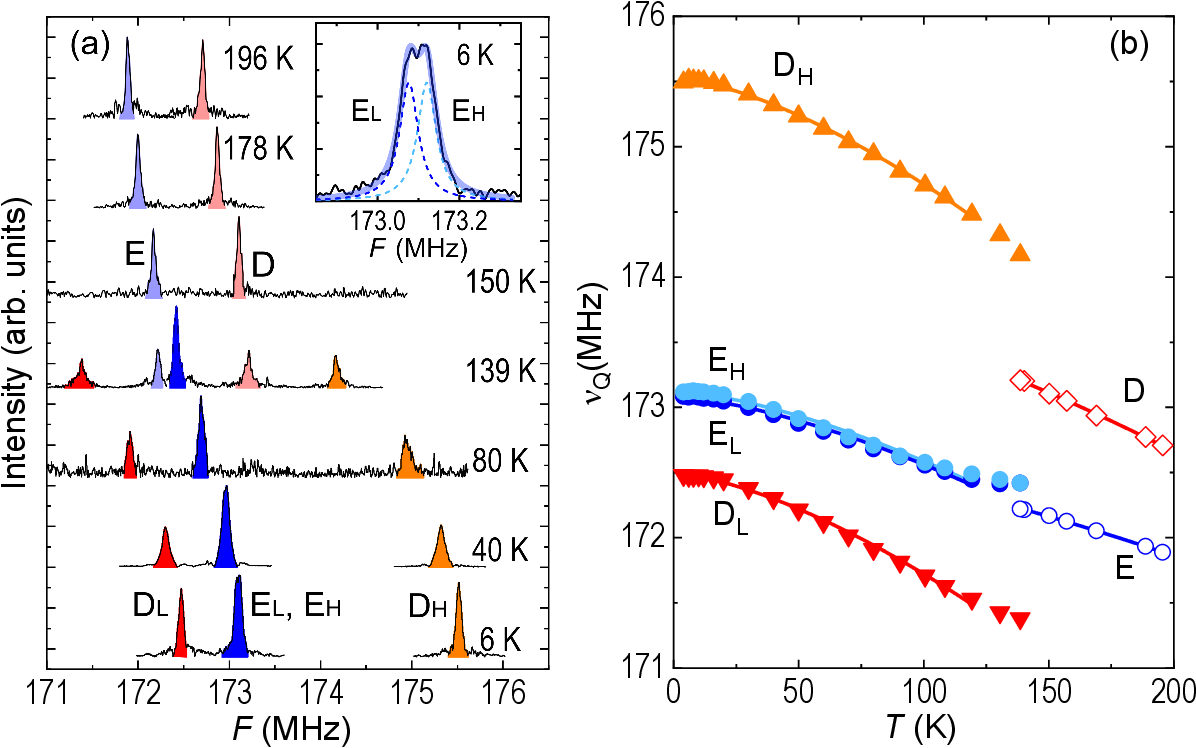}
\end{center}
\caption{
(a) Temperature evolution of the \I-NQR spectra. The inset shows the central peak at $6$~K.  
(b) Temperature dependence of the \I-NQR frequency. The solid lines are the fitting curves of $\nu_{\rm Q}(T)=\nu_{0}(1-\alpha T^{3/2})$.
}
\label{fig2}
\end{figure}

We found two \I-NQR spectral lines at $172.2$ and $173.1$~MHz at $150$~K, as shown in Fig.~\ref{fig2}(a), while sweeping in the range of $170$--$176$~MHz.
Reported \I-NQR studies of I$_3$ anions in quaternary ammonium salts indicated that the observed resonances are ascribed to $\pm 1/2 \leftrightarrow \pm 3/2$ transitions of the terminal iodines~\cite{Yoshioka1983,Bowmaker1968}.
At a terminal iodine, the EFG tensor is nearly axially symmetric \cite{Bowmaker1968} [i.e., $\eta = (V_{xx}-V_{yy})/V_{zz} \simeq 0$, where $V_{ii}$ ($i = x, y, z$) are the diagonal components of the EFG tensor]. 
The NQR frequency, $\nu_{\rm Q}$, is expressed as $\displaystyle \nu_{\rm Q}$ = $\frac{3eQV_{zz}}{20h}\left( 1+\frac{59}{54}\eta^2\right)$, where $e$, $Q$, and $h$ are the elementary charge, the nuclear quadrupole moment of the $^{127}$I nucleus [$-680(10)$~mb \cite{Yakobi2007}], and Planck's constant, respectively. 

In \aI, two crystallographically independent I$_3$ anions, D and E, exist [Fig.~\ref{fig1}(c)]. 
Since the central iodine of each anion is located at the inversion center ($1c$ and $1d$ sites in Wyckoff positions), the terminal iodines of each anion group are equivalent and occupied at 2$i$ sites above $T_\mathrm{CO}$. 
As a result, two independent terminal iodines exist, holding relations I1 = I3 and I2 = I4 (``='' denotes the site equivalency), which agrees with the two distinct lines observed in our NQR spectra.
As discussed below, the signals at the high and low frequencies can be assigned to the D and E sites, respectively, from the calculation of the EFG using first-principles method. 

%%%%%%%%%%%%%%%%%%%%%%%%%%% FIG2 %%%%%%%%%%%%%%%%%%%%%%%%%%%%%%%%%%%
%%%%%%%%%%%%%%%%%%%%%%%%%%% FIG2 %%%%%%%%%%%%%%%%%%%%%%%%%%%%%%%%%%%

At $139$~K, five distinct lines are observed, as shown in Fig.~\ref{fig2}(a), and the three lines remain at the lowest temperature.
The temperature of $139$~K corresponds to $T_{\rm CO}$, and the coexistence of high-temperature and low-temperature peaks in the spectrum indicates that the phase transition is a first-order transition.
Below $T_{\rm CO}$, four lines are expected according to the loss of the inversion center by the CO transition \cite{Kakiuchi2007}.
Although the number of observed lines below $T_{\rm CO}$ is three, the intensities are approximately 1:2:1 from the low-frequency side, suggesting that the NQR spectra consist of four lines.
The central peak below $T_{\rm CO}$ splits slightly, as shown in the inset of Fig.~\ref{fig2}(a). 
We considered that two lines nearly overlap at $\nu_{\rm Q} \approx 173$ MHz because they exhibit nearly the same magnitude of EFG at the terminal iodine sites.

%%%%%%%%%%%%%%%%%%%%%%%%%%% TABLE1 %%%%%%%%%%%%%%%%%%%%%%%%%%%%%%%%%%%
\begin{table}[tbp]
\caption{
Fitted parameters obtained by fitting $\nu_{\rm Q}(T)$ at each site with Eq.~(\ref{nuQ}).}
\begin{ruledtabular}
    \begin{tabular}{cccc} 
        &&$\nu_0$ (MHz)&$\alpha$ (K$^{-3/2}$)\\\hline
        \multirow{3}{*}{D}&$T > T_{\mathrm{CO}}$&173.94(1)&2.61(4)\\
            &\multirow{2}{*}{$T < T_{\mathrm{CO}}$}&172.493(6)&4.47(6)\\
                &&175.528(4)&4.64(4)\\\hline
        \multirow{3}{*}{E}&$T > T_{\mathrm{CO}}$&172.72(1)&1.76(3)\\
            &\multirow{2}{*}{$T < T_{\mathrm{CO}}$}&173.081(7)&3.00(6)\\
                &&173.123(9)&3.09(9)\\
    \end{tabular}
\end{ruledtabular}
\label{table1}
\end{table}
%%%%%%%%%%%%%%%%%%%%%%%%%%% TABLE1 %%%%%%%%%%%%%%%%%%%%%%%%%%%%%%%%%%%

All the $\nu_{\rm Q}$'s increase upon cooling as a result of lattice contraction, as shown in Fig.~\ref{fig2} (b).
The slope of $\nu_{\rm Q}$ at the D site is larger than that at the E site above $T_{\rm CO}$, and below $T_{\rm CO}$, the slopes of $\nu_{\rm Q}$ at the high- and low-frequency signals are larger than that at the central frequency. 
An empirical formula for temperature-dependent $\nu_{\rm Q}(T)$ considering thermal expansion of the unit cell \cite{Christiansen1976}, 
\begin{equation}
    \nu_{\rm Q}(T)=\nu_{0}(1-\alpha T^{3/2})
    \label{nuQ}
\end{equation}
fits the experiments well [solid lines in Fig.~\ref{fig2}(b)], and we summarize the fitted parameters in Table~\ref{table1}.
Comparing $\alpha$ values, $\alpha$ at the D site is $1.5$ times larger than at the E site above $T_{\rm CO}$, and $\alpha$ at the two outer lines is also $1.5$ times larger than at the central line below $T_{\rm CO}$.
The $\alpha$ value that is dependent on the iodine site can be attributed to the distinct principal axes of the EFG at D and E, relative to that of the thermal expansion.
We concluded that the two outer lines originate from the D anion (hereafter, we refer to them as \Dh\ and \Dl), and the overlapped central line originates from the E anion (E$_{\rm H}$ and E$_{\rm L}$).
These results show that $\nu_{\rm Q}$ of the D site changes significantly because of the CO transition, whereas the change in $\nu_{\rm Q}$ of the E site is negligible.

%%%%%%%%%%%%%%%%%%%%%%%%%%% TABLE2 %%%%%%%%%%%%%%%%%%%%%%%%%%%%%%%%%%%
\begin{table*}[tbp]
    \caption{NQR frequencies of experiments (\nuQexp) and calculation (\nuQcal) for terminal iodines of I$_3$ anions.
    }
    \begin{ruledtabular}
    \renewcommand{\arraystretch}{1.25}
    \begin{tabular}{rcclccl}
                                           & \multicolumn{3}{c}{D}                & \multicolumn{3}{c}{E}               \\ \cline{2-4} \cline{5-7}
                                           & \nuQexp (MHz) & \nuQcal (MHz) &  site   & \nuQexp (MHz) & \nuQcal (MHz) &  site \\ \hline
    150 K ($T>T_{\rm CO}$)                 & 173.10       & 174.0        & I1(=I3) & 172.17       & 173.3        & I2(=I4) \\
    \hline
    \multirow{2}{*}{30 K ($T<T_{\rm CO}$)} & 175.40       & 175.2        & I1      & 173.04       & 172.9        & I2      \\
                                           & 172.38       & 173.8        & I3      & 173.00       & 172.6        & I4     
    \end{tabular}
    \end{ruledtabular}
    \label{table2}
    \end{table*}
    %%%%%%%%%%%%%%%%%%%%%%%%%%% TABLE2 %%%%%%%%%%%%%%%%%%%%%%%%%%%%%%%%%%%
We calculated the EFG at each iodine position using the FLAPW method to clarify the relationship between the measured NQR signals and the actual iodine positions~\cite{Supplement}. 
Since the EFG is usually very sensitive to internal atomic coordinates~\cite{Blaha_EFG} and hydrogen atom positions determined from x-ray diffraction were not so accurate~\cite{Edwards_HB2022}, we performed structural optimization for iodine and hydrogen atom positions. 
The relaxed structures with the Crystallographic Information File format and detailed calculation method are included in the Supplemental Material \cite{Supplement}.
The calculated NQR frequencies $\nu_{\rm Q}$'s are summarized in Table~\ref{table2} along with the experimental values. 
At $150$~K, the calculations show that $\nu_{\rm Q} = 174.0$~MHz at the D site, which is $0.4$\% larger than $\nu_{\rm Q} = 173.3$~MHz at the E site.
Experimentally, the difference in frequency between the two NQR lines is approximately $0.5$\%, which is in good quantitative agreement.
Hence, the high- and low-frequency lines are interpreted as signals originating from D and E anions.

At $30$~K, the calculated $\nu_{\rm Q}$ at the D site splits significantly into two lines with higher and lower frequencies, while the change at the E site is small.
To discuss the amount of change in the D and E sites during the CO transition, we define 
\begin{equation}
    \left( \frac{\Delta \nu}{\nu}\right)_{{\rm Q},k} \equiv \frac{(\nu_{{\rm Q},k_H} - \nu_{{\rm Q},k_L})_{30{\rm K}}}{(\nu_{{\rm Q},k})_{150{\rm K}}}
\end{equation}
 where $k$ = D, E sites, resulting in experimental values: $(\Delta \nu/\nu)_{\rm Q,D}$ = $1.75$\% and $(\Delta \nu/\nu)_{\rm Q,E}$ = $0.02$\%; calculated values: $(\Delta \nu/\nu)_{\rm Q,D}$ = $0.74$\% and $(\Delta \nu/\nu)_{\rm Q,E}$ = $0.19$\%. 
The first-principles DFT calculations show that $(\Delta \nu/\nu)_{\rm Q,D}$ is larger than $(\Delta \nu/\nu)_{\rm Q,E}$, which qualitatively explain the experimental results, although $\nu_{\rm Q}$ at the E site still split in the calculations.

%%%%%%%%%%%%%%%%%%%%%%%%%%% FIG3 %%%%%%%%%%%%%%%%%%%%%%%%%%%%%%%%%%%
\begin{figure}[tbp]
    \begin{center}
    \includegraphics[width=8cm]{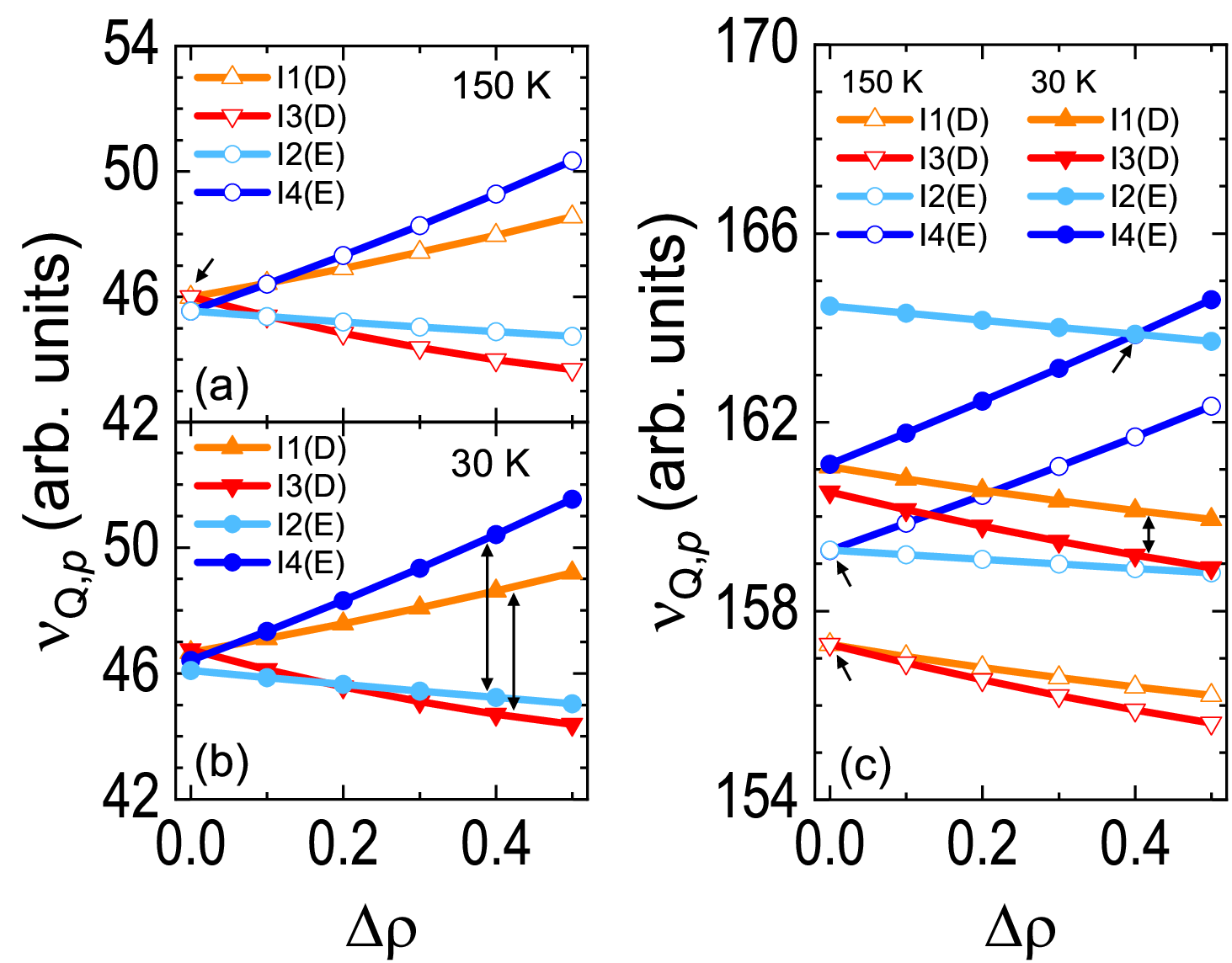}
    \end{center}
    \caption{(a)--(c) $\nu_{{\rm Q}, p}$ calculated by the point charge model as a function of $\Delta \rho$ with $\rho = 0.5 \pm \Delta \rho$ ($0 < \Delta \rho < 0.5$). 
    (a) and (b) show the calculation results based on the $150$ and $30$~K structures, respectively, taking into account only the charge of the ET layers. 
    (c) shows the results for $150$ and $30$~K calculated by incorporating the charge of the I$_3$ layers in addition to the ET layers.
    }
    \label{fig3}
    \end{figure}
%%%%%%%%%%%%%%%%%%%%%%%%%%% FIG3 %%%%%%%%%%%%%%%%%%%%%%%%%%%%%%%%%%%

In order to explore the cause behind the different conduct of the D and E sites, we performed calculations on the EFG that governs $\nu_{\rm Q}$, as a function of the charge-disproportionation ratio in the ET layer, by utilizing the point charge model (see the Supplemental Material \cite{Supplement}) \cite{Gabes1973,Mori1984a}.
Figures~\ref{fig3}(a) and (b) show $\nu_{{\rm Q},p}$ ($p$ = I1--I4) only from the ET layers at $150$~K [Fig.~\ref{fig3}(a)] and $30$~K [Fig.~\ref{fig3}(b)] as a function of $\Delta \rho$ with $\rho = 0.5 \pm \Delta \rho$ ($0 < \Delta \rho < 0.5$).
Here, the positive sign represents rich charge densities $\rho$ for molecules A and B, while the negative sign corresponds to $\rho$ for molecules A$'$ and C in the CO state. 
$\Delta \rho = 0$ is at a high-temperature electronic state, and $\Delta \rho$ = $0.2$--$0.3$ is reported in the CO state.
At $150$~K, because of $\Delta \rho=0$ and the existence of the inversion symmetry, $\nu_{\rm Q,I1}$ and $\nu_{\rm Q,I3}$ ($\nu_{\rm Q,I2}$ and $\nu_{\rm Q,I4}$) must agree. The agreement of them at $\Delta \rho = 0$, indicated by the arrow [Fig.~\ref{fig3}(a)], shows the validity of the calculation. 
There is no clear difference when we compare the $\nu_{\rm Q}$ at $150$ and $30$~K. Moreover, as $\Delta \rho$ increases, the difference within each anion, i.e., $\nu_{\rm Q,I1} - \nu_{\rm Q,I3}$, $\nu_{\rm Q,I2} - \nu_{\rm Q,I4}$, increases [two-headed arrows in Fig.~\ref{fig3}(b)]. This is inconsistent with the experimental results, in which the E line hardly changes in the temperature-dependent spectra and the D line changes significantly.

Next, we computed the EFG incorporating the EFG from the negative charges of the I$_3$ layers and show $\nu_{{\rm Q},p}$ in Fig.~\ref{fig3}(c).
The magnitudes of the change in $\nu_{\rm Q,I2}$ and $\nu_{\rm Q,I4}$ across the CO transition differ significantly: $\nu_{\rm Q,I2} - \nu_{\rm Q,I4} = 0$ at $\Delta \rho = 0$ for $150$~K, and that at $\Delta \rho = 0.4$ for $30$~K. 
On the other hand, $\nu_{\rm Q,I1} - \nu_{\rm Q,I3}$ is minimized at $\Delta \rho = 0$ both for $150$ and $30$~K, and $\nu_{\rm Q,I1} - \nu_{\rm Q,I3}$ at all $\Delta \rho$ for $30$~K is greater than zero. 
The less pronounced differentiation in $\nu_{\rm Q}$ at the E site compared to the D site in the CO state is replicated, indicating the substantial role of the interaction between the anion and cation in reproducing the microscopic spectroscopy.

As the origin of this anion--cation interaction, Alemany \textit{et al}. \cite{Alemany2012} pointed out that the hydrogen bonding is important and that the E anion is more strongly bound to hydrogen atoms than the D anion. 
The presence of hydrogen bonding would enhance the anion-cation interaction. 
The observed larger change in $\nu_{\rm Q}$ at the E site is consistent with this argument.
Hydrogen bonding has been discussed experimentally from a structural analysis \cite{Alemany2012} and also by transport measurements \cite{Ivek2017}. 
In this study, we were able to explore the hydrogen bonding from a microscopic point of view.

\subsection{Relaxation rate}
\begin{figure}[tbp]
\begin{center}
\includegraphics[width=8cm]{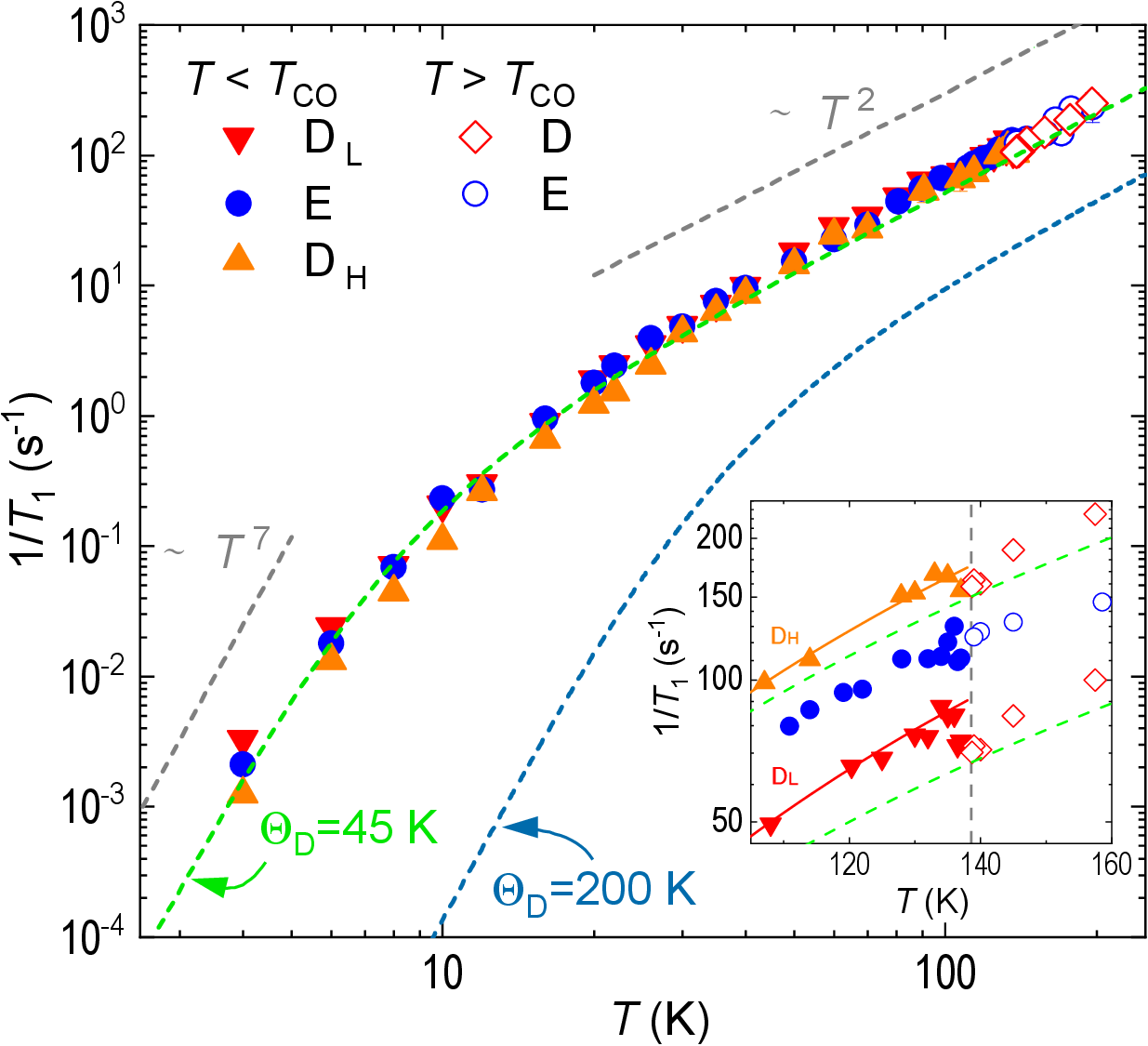}
\end{center}
\caption{
Temperature dependence of $1/T_1$.
The dashed lines are calculated $(1/T_1)_Q$ with $\Theta$ = $45$~K and $200$~K. 
The inset shows the temperature dependence of $1/T_1$ in the vicinity of $T_{\rm CO}$. 
For clarity, the \Dh\ and \Dl\ data are multiplied and divided by $1.5$, respectively.
The solid curves are $1/T_1=(1/T_1)_Q+\mathrm{const.}\times \exp(-\Delta/k_B T)$.
} 
\label{fig4}
\end{figure}

In Fig.~\ref{fig4}, we show the temperature-dependent nuclear spin-lattice relaxation rate $1/T_{1}$ for each spectral line. 
No significant variation in $1/T_1$ for the distinct lines is observed. 
The $1/T_1$ nearly follows power relations in regards to temperature, $1/T_1 \propto T^{\beta}$, which differs from $^{13}$C NMR \cite{Ishikawa2016}. 
The $\beta$ changes from high temperature, $\beta=2$, to low temperature, $\beta \approx 7$, suggesting that the relaxation mechanism is quadrupolar relaxation of nuclei by two-phonon Raman processes~\cite{Iwase2007,Kobayashi2020}.
The $1/T_1$ originating from this process is expressed as \cite{Abragam1961} 
\begin{equation}
  \left( \frac{1}{T_1} \right)_Q\simeq \frac{81\pi}{2}\left( \frac{F_2 \hbar}{mv^2}\right)^2\int_0^{\Omega} \frac{e^{\hbar \omega / k_BT}}{(e^{\hbar \omega / k_B T}-1)^2}\left( \frac{\omega}{\Omega}\right)^6d\omega,\\
  \label{T1}
\end{equation}
where $k_B, m, \Omega$ are the Boltzmann constant, the mass of the $^{127}$I nucleus, and a cutoff frequency, respectively.
$v$ is the sound velocity in the crystal, which is an order of $10^3$ m/s in organic conductors \cite{ImajoSound}. 
$\Omega$ is represented by the temperature $\Theta = (\hbar/k_B)\Omega$ and can be estimated as the Debye temperature, $\Theta = \Theta_D = 200$~K, which is obtained by heat capacity measurements of ET organic conductors.
$F_2$ characterizes the phonon modulation of the EFG and is difficult to determine precisely; however, the simplest estimate, $F_2 = 2\pi\nu_{\rm Q}$, works well~\cite{Klanjvsek2017,Kobayashi2020}. 

The calculated $(1/T_1)_Q$ is plotted as the blue dashed line in Fig.~\ref{fig4} using $\nu_{\rm Q}=173$~MHz
and $\Theta=200$ K, which is significantly smaller than the experimental $1/T_1$. 
Instead, the calculated $(1/T_1)_Q$ using $\Theta =45$~K (green dashed line) reproduces the experiments well. 
This discrepancy suggests that the local $\Theta$ at iodine sites in Eq.~(\ref{T1}) can be smaller than the global Debye temperature $\Theta_D$ estimated from macroscopic measurements. 
We can cite several related experiments. 
The $^{129}$I-M\"{o}ssbauer results of $\beta$-ET$_2$I$_3$ have been explained using the local Debye temperature of $\sim100$~K \cite{Wortmann1992}, which is smaller than $\Theta _D\simeq 200 \ \rm K$ estimated from the heat capacity measurement \cite{Stewart1986}. 
Raman spectra also showed a low-lying molecular vibration mode with $27~{\rm cm^{-1}}$ (= $38 \ {\rm K}$) assigned to the libration of I$_3$ anions, a reciprocating motion with a fixed position in the middle of the I$_3$ ``stick'' (i.e., the central iodine). 
These arguments are consistent with our observations.

%%%%%%%%%%%%%%%%%%%%%%%%%%%%% FIG4 %%%%%%%%%%%%%%%%%%%%%%%%%%%%%%%%%
%%%%%%%%%%%%%%%%%%%%%%%%%%%%% FIG4 %%%%%%%%%%%%%%%%%%%%%%%%%%%%%%%%%

Around the $T_\mathrm{CO}$, we found small peaks for $1/T_1$ of the charge-sensitive D$_\mathrm{H}$ and D$_\mathrm{L}$ lines, as shown in the inset of Fig.~\ref{fig4}. 
Notably, that the peak temperature is located just below $T_\mathrm{CO}$, by which we consider the source for the additional relaxation is not a critical slowing down for the CO. 
A similar peak for $1/T_1$ just below the $T_\mathrm{CO}$ was observed by $^{13}$C-NMR for the charge-rich site~\cite{Ishikawa2016}.
Reference~[\onlinecite{Ishikawa2016}] argues that an emergent antiferromagnetic zigzag chain ($S=1/2$) for $T<T_\mathrm{CO}$ undergoes a singlet state with a sizable gap of $\Delta = 40$~meV~\cite{Tanaka2016}. 
We plot $\displaystyle 1/T_1=(1/T_1)_Q+\mathrm{const.}\times \exp(-\Delta/k_B T)$ as the solid lines in the inset of Fig.~\ref{fig4} and find  good agreements with the experimental $1/T_1$ for the \Dh\ and \Dl\ sites. 
However, we cannot conclude that magnetic fluctuation from charge-ordered ET layers is the source for the peaks of $1/T_1$ because $1/T_1$ for \Dh\ and \Dl\ sites that are close to charge-rich and charge-poor ET molecules show comparable enhancements.

When we consider the difference in the relaxation mechanisms of $^{13}$C and $^{127}$I (i.e., magnetic fluctuation of the local spin density for the former and EFG fluctuation for the latter), we can point out another possibility, other than an emergent spin chain that undergoes a spin-singlet state.
In charge-density-wave (CDW) systems, it is argued that coherence peaklike properties just below the transition temperature can emerge, originating from singularity of the density of states across the CDW gap \cite{Gruner1988, Matus1999, Maniv1982}. 
The electron density is also expected to follow an activated $T$ dependence; then, we cannot discriminate between the energy gaps in the CDW and spin-singlet Mott systems. 
The first-principles calculations also show a similar singularity for the gap opening due to Coulomb repulsion~\cite{Tsumuraya2020,Tsumuraya2021}. 
Detailed theoretical analyses considering the gap structures and the relationships between the magnitudes of the gap and leading electronic/magnetic interactions are necessary to determine the origins of the gap.

In summary, we obtained $^{127}$I-NQR spectra and $1/T_1$ for \aI\ that undergoes a CO transition at ambient pressure. 
The first-principles calculations accurately reproduced the NQR frequency of each spectral line, which enabled us to examine the microscopic interactions. 
The spectral line at the D site splits significantly, while that at the E site is negligibly small. 
These results are qualitatively reproduced by EFG calculations based on a point-charge model that incorporates contributions from both ET and I$_3$ layers, suggesting the importance of the anion-cation interaction.
Furthermore, $1/T_1$ is predominantly determined by a two-phonon process characterized by the local vibration of I$_3$ anions. 
The charge order phase transition is first-order, without a critical slowing down of the electronic charges. 
Just below the $T_\mathrm{CO}$, $1/T_1$ shows an additional relaxation similar to the reported $^{13}$C NMR. 

\begin{acknowledgments}
We are grateful to H.~Sawa and H.~Seo for their helpful discussions.
The computations in this study were mainly conducted using the computer facilities of MASAMUNE at IMR, Tohoku University, Japan.
This work was partially supported by the Japan Society for the Promotion of Science KAKENHI Grants No. 20K14401, No. 21K03438, No. 20K03870, No. 21K03426, and No. 19K21860.
\end{acknowledgments}

% \bibliography{aps}
% 
\end{document}

\title{Supplementary Material:\\
``Role of hydrogen bonding in charge-ordered organic conductor $\alpha$-(BEDT-TTF)$_2$I$_3$ probed by $^{127}$I nuclear quadrupole resonance''}
\author{T.~Kobayashi}
 \email{tkobayashi@mail.saitama-u.ac.jp}
 \affiliation{Graduate School of Science and Engineering, Saitama University, Saitama, 338-8570, Japan}
 \affiliation{Research and Development Bureau, Saitama University, Saitama 338-8570, Japan}
 \affiliation{Meson Science Laboratory, RIKEN, Saitama 351-0198, Japan}
 \author{Y.~Kato}
 \affiliation{Graduate School of Science and Engineering, Saitama University, Saitama, 338-8570, Japan}
\author{H.~Taniguchi}
 \affiliation{Graduate School of Science and Engineering, Saitama University, Saitama, 338-8570, Japan}
\author{T.~Tsumuraya}
\affiliation{Magnesium Research Center, Kumamoto University, 860-8555, Japan}
\author{K.~Hiraki}
 \affiliation{Center for Integrated Science and Humanities, Fukushima Medical University, Fukushima, 960-1295, Japan}
\author{S.~Fujiyama}
 \affiliation{Meson Science Laboratory, RIKEN, Saitama 351-0198, Japan}
\date{\today}
%\pacs{75.25.-j, 75.70.Tj, 71.30.+}
%\keywords{Suggested keywords}%Use showkeys class option if keyword display desired
\maketitle
\beginsupplement

\subsection{1. Calculation of electric field gradients from first principles}
All nuclei with a nuclear spin quantum number $I$ $\geq$ 1 have an electric quadrupole moment $Q$, which originated from the nonspherical charge distribution. 
The nuclear quadrupole interaction between $Q$ and the electric-field gradients (EFG) at the nuclei position determines the coupling constant. 
EFG is very sensitive to the deviation of charge from the spherical part of potentials. 

We performed a first-principles density-functional theory (DFT) calculation to compute EFG in $\alpha$-(BEDT-TTF)$_{2}$I$_{3}$ using the all-electron full-potential linearized augmented plane wave (FLAPW) method~\cite{Wimmer1981, Yu1991}. 
In the EFG calculation, we refer to the experimental crystal structures above and below the charge ordering (CO) transition temperature $T \approx 135$~K.
The EFG is usually very sensitive to internal atomic coordinates.
Therefore, we performed structural optimization for the hydrogen and iodine atom positions using the DFT calculations; 
For the H-atom position, the x-ray diffraction relies on electron density at the scattering sites, and hydrogen atoms have low electron density~\cite{Edwards_HB2022}; C--H bond distances in the structures determined by x-ray diffraction is generally shorter than those optimized with DFT method. 

The low-temperature structure is a triclinic structure with the space group of \textit{P}1, which was measured at $30$~K. 
Above the CO temperature, we used a centrosymmetric structure with the space group of \textit{P}{$\bar{1}$}, measured at $150$~K. 
Those structures were obtained with synchrotron x-ray diffraction \cite{Kitou2021}.  
The CIF (crystallographic information framework) files are deposited in the Cambridge structure database (Nos. 208980 and 208981). 

In Table~\ref{table1}, we list the calculated EFG at the terminal iodine sites. 
The NQR frequencies $\nu_{\rm Q}$s were obtained from these results and $\displaystyle \nu_{\rm Q}$ = $\frac{3eQV_{zz}}{20h}\left( 1+\frac{59}{54}\eta^2\right)$ with $Q=-680(10)$~mb \cite{Yakobi2007}, and are summarized in Table~II of the main text. 

In the present DFT calculations, the Kohn--Sham equations are self-consistently solved using the FLAPW method implemented in the QMD-FLAPW code~\cite{Wimmer1981}. 
The exchange-correlation functional used is the generalized gradient approximation proposed by Perdew, Burke, and Ernzerhof~\cite{Perdew1996}. 
The dimensions of the $\vb*{k}$-point mesh used were $4\times  4 \times 2$ for the self-consistent loop. 
The cutoff energies for the LAPW basis and the potential and density were 20.3 and 213.1 Ry, respectively. We set the muffin-tin (MT) sphere radii as 1.26, 0.75, 1.98, and 2.76 Bohr for C, H, S, and I atoms, respectively. The electronic states up to C $(2s)^{2}$, S $(2p)^{6}$, and I $(4d)^{10}$ were treated as core electrons, which are predominantly confined to the MT spheres. 

%%%%%%%%%%%%%%%%%%%%%%%%%%% TABLE1 %%%%%%%%%%%%%%%%%%%%%%%%%%%%%%%%%%%
\begin{table}[tbp]
\caption{The principal values of the EFG tensor $V_{zz}$ and asymmetry parameter $\eta$.
 }
\begin{ruledtabular}
\renewcommand{\arraystretch}{1.25}
\begin{tabular}{rcclccl}
                                  & \multicolumn{3}{c}{D}           & \multicolumn{3}{c}{E}                                                     \\ \cline{2-4} \cline{5-7} 
                                  & $V_{zz} (10^{21} {\rm V/m^2})$ & $\eta$   &    site  & $V_{zz} (10^{21} {\rm V/m^2})$ & $\eta$   &   site   \\ \hline
150 K ($T>T_{\rm CO}$)                 & 70.533          & 0.036 & I1(=I3) & 70.152                                                  & 0.043 & I2(=I4) \\

\hline
\multirow{2}{*}{30 K ($T<T_{\rm CO}$)} & 70.897          & 0.036 & I1      & 69.981                                                  & 0.042 & I2      \\
                                  & 70.375          & 0.034 & I3      & 69.848                                                  & 0.043 & I4     
\end{tabular}
\end{ruledtabular}
\label{table1}
\end{table}
%%%%%%%%%%%%%%%%%%%%%%%%%%% TABLE1 %%%%%%%%%%%%%%%%%%%%%%%%%%%%%%%%%%%

\subsection{2. Calculation of electric field gradients by point charge model}
The EFGs created by each ion at the iodine positions were calculated by the following equation,
\begin{equation}
\label{S2}
    V_{\alpha \beta} = \frac{\partial^2 V}{\partial \alpha \partial \beta} = \sum_{i}Z_{i}\frac{3x_{\alpha}^{i} x_{\beta}^{i} - \delta_{\alpha \beta}r^2_{i}}{r^5_{i}},
\end{equation}
where $\alpha$ and $\beta$ are $x$, $y$, or $z$, $Z_i$ is the charge on the ion $i$, $x_{\alpha}^i$ ($\alpha$ = $x,y,z$) is the $\alpha$ component of the coordinate of ion $i$ with respect to the calculated iodine position, and $r_i$ is the distance from the iodine position to ion $i$. 
The charge of ET molecule is $+0.5$ in the paramagnetic state, and in the charge-ordered state, its sum of A or B molecules (rich site) and A' or C molecules (poor site) is $+1$. 
The charge determined for each ET molecule in such a way was distributed to each atom in the ratio of the square of the highest occupied molecular orbital coefficient \cite{Mori1984a}. 
For the I$_3^-$ ions, the charges were distributed according to Ref.~\cite{Gabes1973}, excluding only the calculated iodine positions.
The range to be included in the calculation was within $\pm 10$ unit cells in each $x$, $y$, and $z$ direction with respect to the calculated iodine position.

\subsection{3. Relaxation curve}
Nuclear spin-lattice relaxation times are generally produced by magnetic and quadrupole relaxation. The former is determined by the transition probability $W_M$ derived from nuclear spin levels of $\Delta m = \pm 1$ \cite{MacLaughlin1971}, and the latter by the ratio of transition probabilities $W_1$ and $W_2$ derived from $\Delta m = \pm 1, \pm 2$, respectively \cite{Daniel1964}. 
Here, the notation of each transition probability follows the references.
For $I = 5/2$, the relaxation curve can be expressed as 
\begin{equation}
    \label{relaxation}
    \frac{M(\infty)-M(t)}{M(\infty)} = c_+ \exp(-k_{+}Wt) + c_- \exp(-k_{-}Wt) 
\end{equation}
where $M(t)$ and $M(\infty)$ are the nuclear magnetization after the saturation pulse and the equilibrium nuclear magnetization at $t \rightarrow \infty$, respectively.
In the case of magnetic relaxation, the relaxation rate is expressed as $1/T_1 = 2W = 2W_{\rm M}$ and $k_{\pm} = 20,6$.
For quadrupole relaxation, the relaxation rate is $1/T_1 = 2W = 2W_1$ and $k_{\pm}$ is expressed as
\begin{equation}
    k_{\pm} = \frac{14}{25}\left(1+\frac{W_2}{W_1}\right) \pm \frac{1}{25} \sqrt{76-88\left(\frac{W_2}{W_1}\right)+61\left(\frac{W_2}{W_1}\right)^2}
\end{equation}
The relaxation curves for magnetic relaxation only and for quadrupole relaxation only given the ratio $W_2/W_1$ can be calculated, and the several cases of coefficients in Eq.~(\ref{relaxation}) are shown in Table~\ref{table3}. 
Here, the calculations were carried out in the high-temperature limit, with the saturation of $\pm 1/2 \leftrightarrow \pm 3/2$ as the initial condition. 

\begin{figure}[tbp]
\begin{center}
\includegraphics[width=14.5cm]{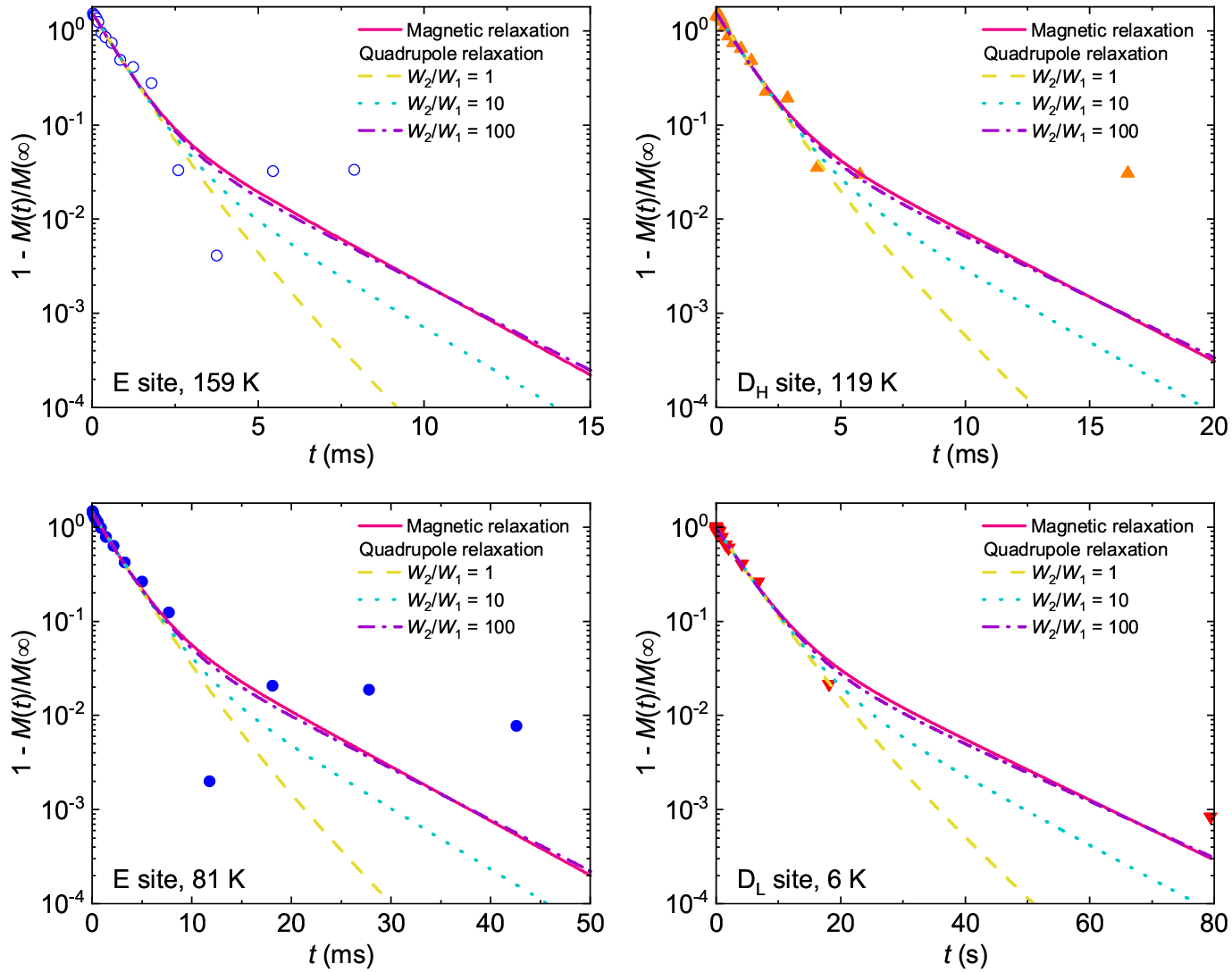}
\end{center}
\caption{
Relaxation profiles of nuclear magnetization at several temperatures and sites. 
Fitting results using Eq.~(\ref{relaxation}) due to magnetic relaxation and quadrupolar relaxation for several $W_2/W_1$ ratios are depicted for different line types.}
\label{figS1}
\end{figure}

Examples of fitting relaxation curves at several sites and temperatures using Eq.~(\ref{relaxation}) are shown in Fig.~\ref{figS1}. 
The recovery profiles do not change with temperature or site. 
As can be seen from the figures, it is also difficult to distinguish which relaxation mechanism, magnetic or quadrupolar, is dominant by the fitting. 
In the high-temperature region, magnetic relaxation might contribute, and no matter which equation is used, there is no influence on the discussion of the temperature dependence of $T_1$. 
Therefore, in this study, $T_1$ was determined using the recovery curve due to magnetic relaxation.
\vskip 1.4\baselineskip

%%%%%%%%%%%%%%%%%%%%%%%%%%% FIGS1 %%%%%%%%%%%%%%%%%%%%%%%%%%%%%%%%%%%
%%%%%%%%%%%%%%%%%%%%%%%%%%% FIGS1 %%%%%%%%%%%%%%%%%%%%%%%%%%%%%%%%%%%

%%%%%%%%%%%%%%%%%%%%%%%%%%% TABLE3 %%%%%%%%%%%%%%%%%%%%%%%%%%%%%%%%%%%
\begin{table}[tbp]
\caption{Coefficients of magnetic and quadrupolar relaxation curves in Eq.~(\ref{relaxation}). 
 }
%\begin{threeparttable}
\begin{ruledtabular}
\renewcommand{\arraystretch}{1.25}
\begin{tabular}{ccccc}
                 & $c_+$          & $c_-$ & $k_+$ & $k_-$                  \\
\hline
$W_M$                     & 0.89286 (=25/28)   & 0.10714 (=3/28)   & 20         & 6        \\
$W_2 / W_1$ = 1     & 0.89286   & 0.10714   & 1.4      & 0.84       \\
$W_2 / W_1$ = 10    & 0.93972   & 0.06028   & 9.0794    & 3.24906   \\
$W_2 / W_1$ = 100   & 0.91848   & 0.08152   & 87.5768    & 25.5432   \\
\end{tabular}
\end{ruledtabular}
%\end{threeparttable}
\label{table3}
\end{table}
%%%%%%%%%%%%%%%%%%%%%%%%%%% TABLE3 %%%%%%%%%%%%%%%%%%%%%%%%%%%%%%%%%%%

% \bibliography{aps}
%    